\documentclass[twocolumn,aps,prb,10pt]{revtex4-1}
\usepackage{graphicx}
\usepackage{color}

\begin{document}
\title{Noncovalent Interactions by QMC: Speedup by One-Particle Basis-Set Size Reduction}
\author{Mat\'u\v{s} Dubeck\'{y}}
\email{matus.dubecky@osu.cz}
\affiliation{Department of Physics, Faculty of Science, University of Ostrava, 30. dubna 22, 701 03 Ostrava, Czech Republic}
\affiliation{RCPTM, Department of Physical Chemistry, Faculty of Science, Palack\'y University
Olomouc, t\v{r}.~17~listopadu 12, 771 46 Olomouc, Czech Republic}

\date{\today}

\begin{abstract}
While it is empirically accepted that the fixed-node diffusion Monte-Carlo (FN-DMC) depends only weakly on the size (beyond a certain reasonable level) of the one-particle basis sets used to expand its guiding functions, limits of this observation are not settled yet. Our recent work indicates that under the FN error cancellation conditions, augmented triple zeta basis sets are sufficient to achieve high-quality  benchmark single-point energy differences in a number of small noncovalent complexes. In this preliminary progress report, we report on a possibility of significant truncation of the one-particle basis sets used to express the FN-DMC guiding functions, that has no visible effect on the accuracy of the production energy differences. The proposed scheme shows negligible increase of the local energy variance, indicating that the total CPU cost of large-scale benchmark noncovalent interaction energy FN-DMC calculations employing Gaussians may be reduced.
\end{abstract}
\maketitle

In the domain of benchmark ab-initio noncovalent interaction energy calculations, fixed-node diffusion Monte Carlo (FN-DMC) method provides a promising alternative to the commonly used coupled-cluster (CC) approaches like CCSD(T)\cite{Dubecky2016rev}. Accurate FN-DMC interaction energies (to 0.1 kcal/mol vs. CCSD(T)/CBS) have been recently reported on a number of small/medium noncovalent closed-shell complexes\cite{Gurtubay2007,Santra2008,Ma2009,Korth2011,Gillan2012,Dubecky2013,Benali2014,Hamdani2014}. In addition to accuracy, FN-DMC is attractive also for its favorable low-order polynomial CPU cost scaling\cite{Foulkes2001rev,Bajdich2009rev,Austin2012rev} and favorable FN error cancellation\cite{Mella2003,Diedrich2005,Dubecky2014} that enabled its use in medium/large complexes\cite{Sorella2007,Korth2008,Santra2008,Ambrosetti2014,Benali2014,Hamdani2014} where CC methods were intractable (in original formulation and reasonable basis set) until recently\cite{Yang2014}. Furthermore, direct treatment of  
extended\cite{Raza2011,Santra2011,Hongo2015,Karalti2012,Shulenburger2013,Cox2014,Quigley2014,Shulenburger2014,Hamdani2014,Mostaani2015} and/or multireference systems\cite{Horvathova2013} makes FN-DMC an attractive many-body method worth of further research and development.

While it is empirically accepted that the FN-DMC results depend only weakly on the one-particle basis sets\cite{Dubecky2016rev} used to expand the FN-DMC guiding functions, limits of this assumption in energy differences remain unclear. Our recent work e.g. indicates that in FN-DMC calculations using DFT-based single-determinant guiding functions, the basis set cardinality is not as important as the presence of augmentation functions\cite{Dubecky2014}.
~
An example of ammonia dimer complex well illustrates this behavior: a sequence of VTZ, VQZ and aug-VTZ basis sets generates the interaction energies of \mbox{-3.33$\pm0.07$}, \mbox{-3.47$\pm0.07$} and \mbox{-3.10$\pm0.06$} kcal/mol\cite{Dubecky2014}, while the complete-basis-set(CBS) CCSD(T) benchmark  value at the same geometry amounts to \mbox{-3.15 kcal/mol}\cite{Jurecka2006,Takatani2010}. Augmented triple zeta basis sets were confirmed to be sufficient to achieve a level of 0.1 kcal/mol in the final FN-DMC interaction energies in numerous cases\cite{Dubecky2013,Dubecky2014}. Some residual errors that remain in certain types of noncovalent interactions (e.g. stacking or hydrogen bonds combined with multiple bonding) are not yet understood and require further attention\cite{Dubecky2014rev}. 

QMC is nevertheless a very promising methodology, and its main limitation in area of noncovalent interactions is the CPU cost that stems from the tight statistical convergence required in case of small energy differences. It is therefore important to map out strategies of possible CPU cost reduction. For instance, presence of the FN error cancellation\cite{Mella2003,Diedrich2005,Dubecky2013,Dubecky2014rev,Dubecky2016rev} enabled use of economic Jastrow factor with electron-electron and electron-nucleus terms instead of the more demanding one including also electron-electron-nucleus terms\cite{Dubecky2014}.

Here we report on a possibility of truncation of the one-particle Gaussian basis sets used to expand the orbitals in the single Slater determinant FN-DMC guiding functions without affecting the accuracy of the final energy differences. A series of tests\cite{tbp} led us to the finding that high angular momentum basis functions are not critically important\cite{Diedrich2005,Xu2013} and the augmentation basis set is necessary, but the acceptable accuracy is achieved already with a single diffuse $s$ function per atom used instead of the common aug- set of basis functions\cite{tbp}.

In the following, we compare the two types of basis sets used to express the orbitals in single-determinant FN-DMC guiding functions. The results obtained with trimmed basis set (e.g. [3s3p2d]+[1s] on carbon atom) denoted as s-rVTZ are compared to the ones obtained with the standard aug-VTZ basis set ([3s3p2d1f]+[1s1p1d1f] on carbon atom). Surprisingly, the (single-point) interaction energies produced with the trimmed bases reveal no statistically significant differences with respect to the reference calculations, and, no significant increase of the local energy variance, implying that the equivalent-quality FN-DMC results are available at costs lower than assumed to date.

The considered test set contains seven noncovalent closed-shell complexes (with geometries obtained) from the A24 database\cite{Rezac2013}: ammonia dimer (AM...AM), water-ammonia (WA...AM), water dimer (WA...WA), ammonia methane complex (AM...MT), methane dimer (MT...MT), hydrogen fluoride dimer (HF...HF), and HCN dimer (HCN...HCN). The atomic cores in these complexes were replaced by the effective core potentials (ECPs) developed by Burkatzki et al.\cite{Burkatzki2007,ClaudiaPC}. The ECPs were used in combination with Dunning-type aug-VTZ basis sets or their truncated counterparts (s-rVTZ, see above). Single-determinant Slater-Jastrow\cite{Bajdich2009rev} trial wave functions were constructed using orbitals from B3LYP ({\tt GAMESS}\cite{gamess} code) and
the Schmidt-Moskowitz\cite{Moskowitz1992} isotropic Jastrow factors\cite{Bajdich2009rev} including electron-electron and electron-nucleus terms\cite{Dubecky2014} were expanded in a fixed basis set of polynomial Pad\'{e} functions\cite{Bajdich2009rev}. The Jastrow parameters were refined by the Hessian driven variational Monte Carlo optimization using at least 10x10 iterations and linear combination\cite{Umrigar2005} of energy (95\%) and variance (5\%) as a cost function. The orbital parameters were frozen to keep the nodes of the guiding functions intact.  The production FN-DMC runs used time step of 0.005~a.u and the  \mbox{T-moves} scheme for the treatment of ECPs\cite{Casula2006}. The target walker populations in FN-DMC amounted to 16-32k. All QMC calculations were performed using the code {\tt QWalk}\cite{Wagner2009}.

%%%%%%%%%%%
\begin{table*}[ht!]
\caption{Comparison of interaction energies $\Delta E$ (kcal/mol) and local energy variances in dimers $\sigma^2$ (a.u.$^2$) from FN-DMC calculations using aug-VTZ ($\Delta E$s taken from Ref.~\citenum{Dubecky2014}) and trimmed, s-rVTZ, basis sets, the related counts of the basis functions $M$ in the dimer of the given complex, and, ideal expected speedup ($s_i$, see text) of s-rVTZ vs. aug-VTZ calculation.}
\begin{tabular}{l|ccc|ccc|c}
\hline
\hline
          & \multicolumn{3}{c|}{aug-VTZ}    & \multicolumn{3}{c|}{s-rVTZ}      \\
Complex   & $\Delta E$      &  $\sigma^2$ & $M$  & $\Delta E$        & $\sigma^2$ & $M$  & $s_i$    \\
\hline
AM...AM   & -3.30$\pm$0.04  & 0.433 & 228~~   & -3.36$\pm$0.08 & 0.445 & 106  & 2.09 \\
AM...WA   & -6.71$\pm$0.07  & 0.561 & 205~~   & -6.64$\pm$0.09 & 0.578 & 96   & 2.07 \\
WA...WA   & -5.30$\pm$0.05  & 0.667 & 182~~   & -5.25$\pm$0.09 & 0.682 & 86   & 2.07 \\
HF...HF   & -4.89$\pm$0.05  & 0.960 & 136~~   & -4.89$\pm$0.10 & 0.968 & 66   & 2.04 \\
AM...MT   & -0.83$\pm$0.06  & 0.360 & 251~~   & -0.77$\pm$0.07 & 0.364 & 106  & 2.34 \\
MT...MT   & -0.63$\pm$0.03  & 0.271 & 274~~   & -0.65$\pm$0.04 & 0.282 & 126  & 2.09 \\
HCN...HCN~& -5.09$\pm$0.08  & 0.582 & 226~~   & -4.97$\pm$0.08 & 0.605 & 112  & 1.94 \\
\hline\hline
%$^a$~\cite{Dubecky2014}\\
\end{tabular}
\label{tabres}
\end{table*}
%%%%%%%%%%%%%%%%%

The FN-DMC results obtained with single-determinant Slater-Jastrow guiding functions expanded in aug-VTZ and reduced s-rVTZ basis sets are reported in Table~\ref{tabres}. The interaction energies obtained with \mbox{s-rVTZ} basis are clearly compatible with the \mbox{aug-VTZ} data in {\em all} considered cases (Figure~1), that is a non-trivial and very interesting observation, since, in the trimmed case, the total number of basis functions is reduced by more than two times, and, the local energy variance increases only by up to 4\% (Table~\ref{tabres}). 

Why is this the case? As mentioned above, VTZ or even VQZ bases do not lead to the correct energy differences in the similar setting, so the non-augmented rVTZ basis sets could hardly lead to the more accurate results. We thus presumably attribute the observed agreement of s-rVTZ and aug-VTZ results to the similar degree of FN error cancellation and sufficient sampling of interstitial regions secured by the single $s$ function per atom, working even in such a complex like HF dimer\cite{tbp}. The modest increase of $\sigma^2$ is attributed to the use of identical occupied-shell contractions.

\begin{figure}[ht!]
 \centering
 \includegraphics[width=200pt]{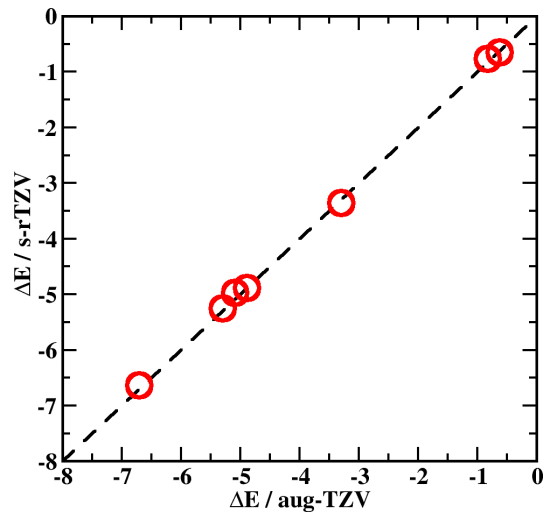}
 % graf.png: 0x0 pixel, 300dpi, 0.00x0.00 cm, bb=
 \caption{Demonstration of the agreement between the FN-DMC interaction energies $\Delta E$ (kcal/mol) from Table~\ref{tabres}. calculated with aug-VTZ and truncated s-rVTZ basis sets. The error bars (not shown) are smaller than the symbol size. }
    \label{fig}
\end{figure}

Let us discuss the CPU cost benefits of the proposed basis-set size trimming. The Slater matrix evaluation step scales as $O(MN^3)$ where $M$ is the number of basis functions and $N$ is the number of electrons. The ideal speedup with reduced $M$ for asymptotically large fixed $N$ is proportional to 
\begin{equation}
s_i=\frac{\sigma^2_{\mathrm{ref}}M_{\mathrm{ref}}}{\sigma^2_{\mathrm{trim}}M_{\mathrm{trim}}}
\end{equation}
where $M_{\mathrm{ref}}$ is the number of basis functions in the reference basis set (aug-VTZ), $M_{\mathrm{trim}}$ is the count of the basis functions in the trimmed case (s-rVTZ), and $\sigma^2$s denote the respective local energy variances.
As the formula (1) indicates, the cost savings start to become interesting for $M_{\mathrm{trim}}$ significantly smaller than $M_{\mathrm{ref}}$  and only if the ratio of local energy variances $\sigma^2_{\mathrm{ref}}/\sigma^2_{\mathrm{trim}}$ does not outweight the gain from $M_{\mathrm{ref}}/M_{\mathrm{trim}}$. Furthermore, the observed speedup is smaller than the theoretical limit $s_i$, since, other routines, like evaluation of Jastrow factor, pseudopotentials and/or distance matrix updates, also add to the overall cost. The observed speedup thus approaches $s_i$ only for sufficiently large $M$ where the Slater matrix updates dominate the overall CPU cost. Even in the theoretical limit, the CPU cost gain grows linearly that may appear unimportant. Nevertheless, the FN-DMC calculations are very expensive in general, and frequently millions of CPU-hours are invested in valuable projects (e.g. Ref.~\citenum{Ambrosetti2014}). Therefore, in our opinion, {\it any} CPU cost reduction is important, and the scheme 
presented here (i.e. use of s-rVTZ instead of aug-VTZ, in combination with Jastrow factor containing electron-electron and electron-nucleus terms) asymptotically offers about two-fold speedup (cf. Table~\ref{tabres} and examples below).

The following examples illustrate the wave function evaluation speedups (compare to $M_{\mathrm{ref}}$/$M_{\mathrm{trim}}$) in pure Slater and Slater-Jastrow runs:

i) In the HCN dimer (20 electrons) where  $M_{\mathrm{ref}}$/$M_{\mathrm{trim}}=2.02$, the practical Slater evaluations with s-rVTZ are 1.5 times faster than in the case of aug-VTZ and the Slater-Jastrow run achieves an evaluation cost speedup of 1.22, rather far from the expected value. This indicates that the system is ``small'' and routines other than the Slater matrix value and Laplacian updates  are still important.

ii) In the case of a larger complex, coronene...H$_2$ with 110 electrons ($M_{\mathrm{ref}}$/$M_{\mathrm{trim}}=1402/692=2.03$), the speedup of a pure Slater run is 1.92-fold while the FN-DMC calculation using Slater-Jastrow guiding function achieves a speedup factor of 1.74. Note that the local energy variance ratio $\sigma^2_{\mathrm{ref}}/\sigma^2_{\mathrm{trim}}=2.590
/2.707=0.96$ so the total observed speedup of the calculation achieves a factor of 1.67 ($s_i$=1.94). In larger systems, the speedups clearly grow toward the theoretical limit and become interesting.

In conclusion, a new one-particle basis set truncation scheme has been shown to considerably reduce computational requirements while retaining a full accuracy in energy differences of seven noncovalent complexes. This finding presumably extends the range of applicability for FN-DMC method in area of noncovalent interactions.
Further tests including extended sets of molecules, other truncation schemes, and various combinations of tuned bases with various Jastrows are now underway\cite{tbp}.

\scriptsize
\acknowledgements
The author is grateful to Claudia Filippi for sharing an 
improved ECP for H and to Lubos Mitas, Petr Jure\v{c}ka, and Ren\'e Derian for stimulating and fruitful discussions. 
Financial support from the University of Ostrava (IRP201558), Ministry of Education, Youth and Sports of the Czech Republic (project LO1305), VEGA (VEGA-2/0130/15), and, computational resources provided by the Ministry of Education, Youth and Sports of the Czech Republic under the Projects CESNET (LM2015042) and CERIT-Scientific Cloud (LM2015085) provided within the program Projects of Large Research, Development and Innovations Infrastructures, are gratefully acknowledged.

%\bibliographystyle{unsrt}
%\bibliography{rest}

\end{document}